\documentclass[preprintnumbers,amsmath,11pt,amssymb,floatfix,superscriptaddress,nofootinbib]{revtex4}
\usepackage{graphicx}
\usepackage{epsfig}
\usepackage{bm}
\usepackage{amsfonts}

\input epsf

\begin{document}

\title{\Large On the stability of the dark energy based on generalized uncertainty principle}

\author{Antonio Pasqua}
\email{toto.pasqua@gmail.com} \affiliation{Department of Physics,
University of Trieste, Via Valerio, 2 34127 Trieste, Italy.}

\author{Surajit Chattopadhyay}
\email{surajit_2008@yahoo.co.in, surajcha@iucaa.ernet.in}
\affiliation{ Pailan College of Management and Technology, Bengal
Pailan Park, Kolkata-700 104, India.}

\author{Iuliia Khomenko}
\email{ju.khomenko@gmail.com} \affiliation{Heat-and-Power Engineering Department National Technical
University of Ukraine "Kyiv Politechnical Institute", Kyiv, Ukraine.}

\date{\today}

\begin{abstract}
The new agegraphic Dark Energy (NADE) model (based on generalized uncertainty
principle) interacting with Dark Matter (DM) is considered in this study
via power-law form of the scale factor $a\left(t\right)$. The equation of state (EoS)
parameter $\omega_G$ is observed to have a phantom-like behaviour. The stability of this model is
investigated through the squared speed of sound $v_s^2$: it is found
that $v_s^2$ always stays at negative level, which
indicates instability of the considered model.
\end{abstract}


\maketitle

\section{Introduction}
It is widely accepted by the scientific community that the Universe is undergoing a phase of accelerated expansion, well established by numerous cosmological observations \cite{perl,ast,benn,sperg,teg,avz,allen}.
The responsible for this accelerated expansion is thought to be a missing dark component with negative pressure dubbed as Dark Energy (DE).
Combined analysis of different cosmological observations indicate that about 73$\%$ of the universe consists of
DE, while the other dark component, i.e. Dark Matter (DM), represents about the 24$\%$ of the universe, with the remaining part made by baryonic matter \cite{peiris,ben2}. Many different DE models have been proposed to explain this accelerated expansion, even if most of them cannot explain properly all the features of universe. The dynamical nature of DE can be originated by various fields, although a complete description needs a deeper understanding of the underlying theory of Quantum Gravity (QG).\\
An interesting attempt made in order to better understand the nature of DE is the proposal of the so-called agegraphic
DE (ADE). This model was proposed by Cai \cite{2007cai} to explain the accelerated expansion of the
universe within the framework of quantum gravity. The ADE model assumes that the
observed DE comes from the space-time and matter field fluctuations present in the universe.
The original ADE has the following expression for the energy density $\rho_D$:
\begin{equation}
\rho_{D}=\frac{3n^{2}m_{p}^{2}}{T ^2}, \label{1-1}
\end{equation}
where $n^2$ is a constant, $m_p = \left(8\pi G\right)^{-1/2}$ (with $G$ representing the gravitational constant) is the reduced Planck mass and $T$ is the age of the universe, defined as:
\begin{equation}
T = \int dt = \int \frac{da}{aH}, \label{2-1}
\end{equation}
where $H$ represents the Hubble parameter and $a$ is a scale factor which is function of the cosmic time $t$.\\
The original ADE model suffers from the difficulty to describe appropriately the matter-dominated epoch,
for this reason a new version of ADE was subsequently proposed by Wei and Cai \cite{2008wei}: the time scale they choose for their model was
the conformal time $\eta$  instead of the age of the universe $T$, yielding the new-agegraphic DE
(NADE) model.  The energy density $\rho_D$ of the NADE is given by:
\begin{equation}
\rho_{D}=\frac{3n^{2}m_{p}^{2}}{\eta ^2}, \label{1}
\end{equation}
where $\eta$ is the conformal time, which is defined as:
\begin{equation}
\eta=  \int\frac{dt}{a} = \int \frac{da}{a^2H}, \label{2}
\end{equation}

The NADE model has some new features respect to the original ADE and, for this reason, it is able to overcome some
unsatisfactory points which are present in the ADE model.\\
The agegraphic models of DE have been also investigated in ample details (see \cite{she1,she2,she3} and references therein).\\
An important quantity studied in order to check the stability of a given model, namely
the squared speed of sound $v_{s}^{2}$, is defined as:
\begin{eqnarray}
v_{s}^{2}=\frac{\dot{p}}{\dot{\rho}}
\end{eqnarray}
The sign of $v_{s}^{2}$ is important for stability of a background
evolution. A negative value implies a classical instability of a
given perturbation in general relativity. Myung
\cite{myung2} has observed that $v_{s}^{2}$ for HDE is always
negative for the future event horizon as IR cut-off, while for
Chaplygin gas and tachyon, it is observed to be non-negative. Kim et al.
\cite{kim} found that the squared speed of sound for ADE
is always negative leading to the instability of the perfect fluid
for the model. Moreover, it was found that the ghost QCD DE
model is unstable \cite{20g}. Recently, Sharif and Jawad \cite{sharif} have
shown that interacting new HDE is characterized by negative
$v_{s}^{2}$.\\
A great interest has been recently poned to the Generalized Uncertainty Principle (GUP) and its
consequences \cite{garay,scardi,chang,set1,medved,nozari,har,ling,myung,park} since the Heisenberg uncertainty principle is not expected to be satisfied anymore when quantum gravitational effects
become more important.\\
Even though the GUP has its origins in the string theory \cite{gross,amati,koni}, the GUP provides the minimal length scale, i.e. the Planck scale $l_p$, and it may play a role of evolution of the universe. Especially, we expect that this can modify significantly the evolution of early universe at the Planck scale and inflation.\\
The purpose of this paper is to study the NADE in the light of GUP and it is organized as follow. In Section 2, we describe the NADE model in the framework of GUP. In Section 3, we derive the EoS parameter $\omega_G$ and the squared speed of the sound $v_s^2$ for a particular choice of the scale factor $a\left(t\right)$. Finally, in Section 4 we write the Conclusions of this work.

\section{NADE with Generalized Uncertainty Principle (GUP)}
In this Section, we want to study the main characteristics of the NADE in the light of the GUP.\\
For scale factor $a\left(t\right)$, we consider the following expression of the scale factor \cite{Setare1}:
\begin{equation}
a(t)=a_{0}(t_{s}-t)^{n}, \label{3}
\end{equation}
where $a_{0}$ is the present value of the scale factor and $t_{s}$ and $n$ are  constants. With the choice of the scale factor given in Eq. (\ref{3}), the Hubble parameter $H$ becomes:
\begin{equation}
 H = \frac{\dot{a}}{a} = -\frac{n}{t_{s}-t}. \label{4}
\end{equation}
The metric of a spatially flat, homogeneous and isotropic universe in Friedmann-Lemaitre-Robertson-Walker (FLRW) model is given by:
\begin{equation}
ds^{2}=dt^{2}-a^{2}(t)\left[dr^{2}+r^{2}(d\theta^{2}+sin^{2}\theta d\phi^{2})\right], \label{5}
\end{equation}
where $t$ is the cosmic time, $r$ is referred to the radial component and $\left(\theta , \phi \right)$ are the angular coordinates.\\
The Einstein field equations are given by:
\begin{eqnarray}
H^{2}&=&\frac{1}{3}\rho, \label{6}\\
\dot{H}&=&-\frac{1}{2}(\rho+p),\label{7}
\end{eqnarray}
where $\rho$ and $p$ are, respectively, the energy density and the isotropic pressure (choosing $8\pi G=c=1$).\\
We now consider the interaction between the NADE and DM using GUP.
We start with extending the time-energy uncertainty to the GUP:
\begin{equation}
\Delta E \Delta t > 1 + \zeta \left( \Delta E \right)^2, \label{9}
\end{equation}
in the units of $c = \hbar = k_B =1$.
The parameter $\zeta$ has the Planck length scale $\left( \zeta \approx l_p^2 \approx \ m_p^2  \right)$ . Solving the saturation of the GUP leads to:
\begin{equation}
\Delta E_{G}=  \frac{1}{\Delta t}+\frac{\zeta}{\Delta t^{3}}  =\frac{1}{t}+\frac{\zeta}{t^{3}}, \label{10}
\end{equation}
where we use the relation of $\Delta t  \approx  t$ for cosmological purpose.
 According to the GUP, the energy density $\rho_G$ is defined as \cite{Kim}:
\begin{equation}
\rho_{G}=\frac{\Delta E_{G}}{(\delta t)^{3}}, \label{11}
\end{equation}
where $\delta t$ is given by the K$\acute{a}$rolyh$\acute{a}$zy relation of time fluctuations as follow:
\begin{equation}
\delta t=t_{p}^{2/3}t^{1/3}. \label{11-1}
\end{equation}
$\zeta$ and $t_{p}$ are defined, respectively, as:
\begin{eqnarray}
\zeta &=& \left(\frac{\xi}{n}\right)^{2}, \label{12}\\
t_{p}^{2} &=& \frac{1}{3n^{2}m_{p}^{2}}. \label{13}
\end{eqnarray}
 Consequently, the DE density is described with the two
parameters $(n, \xi)$ as follow:
\begin{equation}
\rho_{G}=\frac{3n^{2}m_{p}^{2}}{t^{2}}+\frac{3\xi^{2}}{t^{4}}. \label{14}
\end{equation}
Substituting $t$ with $\eta$ in Eq. (\ref{14}), the DE energy density based on GUP takes the following form (Kim et al, 2008):
\begin{equation}
\rho_{G}=\frac{3n^{2}m_{p}^{2}}{\eta^{2}}+\frac{3\xi^{2}}{\eta^{4}}. \label{15}
\end{equation}

\section{Interacting DE}
We are now considering interaction between DM and the DE. \\
In order the local energy-momentum conservation law is preserved, the total energy density $\rho_{total}=\rho_G + \rho_m$ must satisfy the following continuity equation:
\begin{equation}
\dot{\rho}_{total}+3H(\rho_{total}+p_{total})=0, \label{15-1}
\end{equation}
where:
\begin{eqnarray}
\rho_{total} &=& \rho_{G}+\rho_{m}, \label{16}\\
p_{total} &=& p_{G}+p_{m}.  \label{17}
\end{eqnarray}
$p_{G}$, $p_{m}$ and $\rho_{m}$ denote, respectively, the pressure of the
GUP based DE, the pressure of DM and the density of the DM.\\
Since in this paper we consider interaction between DE and DM, the two energy densities $\rho_G$ and $\rho_m$ are preserved separately:
\begin{eqnarray}
\dot{\rho}_{G}&+&3H(\rho_{G}+p_{G})=Q, \label{17-1}\\
\dot{\rho}_{m}&+&3H\rho_{m}=-Q, \label{17-2}
\end{eqnarray}
where $Q$ represents an interaction term which can be an arbitrary function of cosmological
parameters. Many candidates have been proposed in order to describe $Q$: the one chosen in this paper is $Q=3H\delta\rho_{m}$, where $\delta$ represents the interaction parameter, which exact value is still under debate.\\
With the choice of $a\left(t\right)$ we have done in Eq. (\ref{3}), the conformal time $\eta$ assumes the following expression:
\begin{equation}
\eta=\frac{(t_{s}-t)^{-n+1}}{a_{0}(n-1)}. \label{18}
\end{equation}
Using the expression of $\eta$ given in Eq. (\ref{18}) in Eq. (\ref{15}), we get the energy density $\rho_{G}$ as a function of cosmic time $t$ as follow:
\begin{equation}
\rho_{G}=3a_{0}^{2}\left[a_{0}^{2}(n-1)^{4}(t_{s}-t)^{4(n-1)}\xi^{2}+n^{2}(n-1)^{2}(t_{s}-t)^{2(n-1)}m_{p}^{2}\right]. \label{19}
\end{equation}
Inserting Eq. (\ref{19}) in the continuity equation for $\rho_G$, we obtain the following expression for $p_G$:
\begin{equation}
\begin{array}{c}
p_{G}=-\frac{a_{0}^{4}(n-1)^{4}(7n-4)(t_{s}-t)^{4(n-1)}\xi^{2}}{n}+\frac{(t_{s}-t)^{-3n}(a_{0}(t_{s}-t)^{n})^{-3\delta}\delta\rho_{m0}}{a_{0}^{3}}\\-a_{0}^{2}(n-1)^{2}n(5n-2)(t_{s}-t)^{2(n-1)}m_{p}^{2}. \label{20}
\end{array}
\end{equation}
Considering Eqs. (\ref{19}) and (\ref{20}), we derive the expression of the EoS parameter $\omega_G$ as follow:
\begin{equation}
\omega_{G}=\frac{p_{G}}{\rho_{G}}. \label{21}
\end{equation}
The behavior of the EoS parameter given in Eq. (\ref{21}) is plotted in Figure 1.
It is clear from Figure 1 that $\omega_G$ stays below $-1$ and, hence, it may be interpreted
that the $\omega_F$ has a phantom-like behavior.
\begin{figure}[h]
\begin{minipage}{16pc}
\includegraphics[width=16pc]{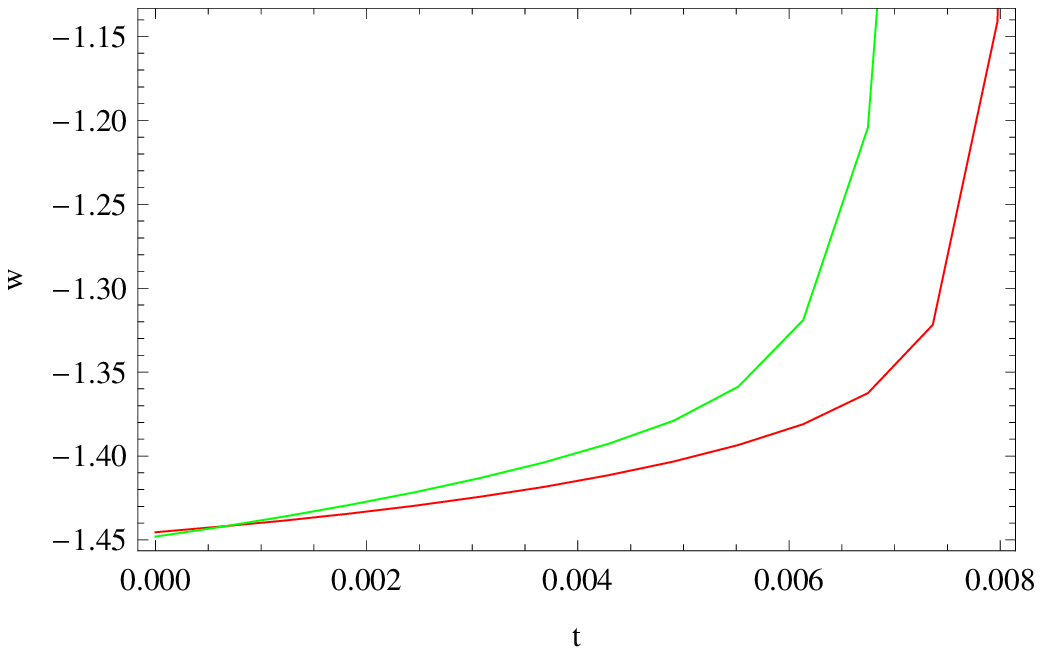}
\caption{\label{label} Evolution of the EoS parameter $\omega_G$ given in Eq. (\ref{21}) against the cosmic
time $t$.}
\end{minipage}\hspace{3pc}%
\begin{minipage}{16pc}
\includegraphics[width=16pc]{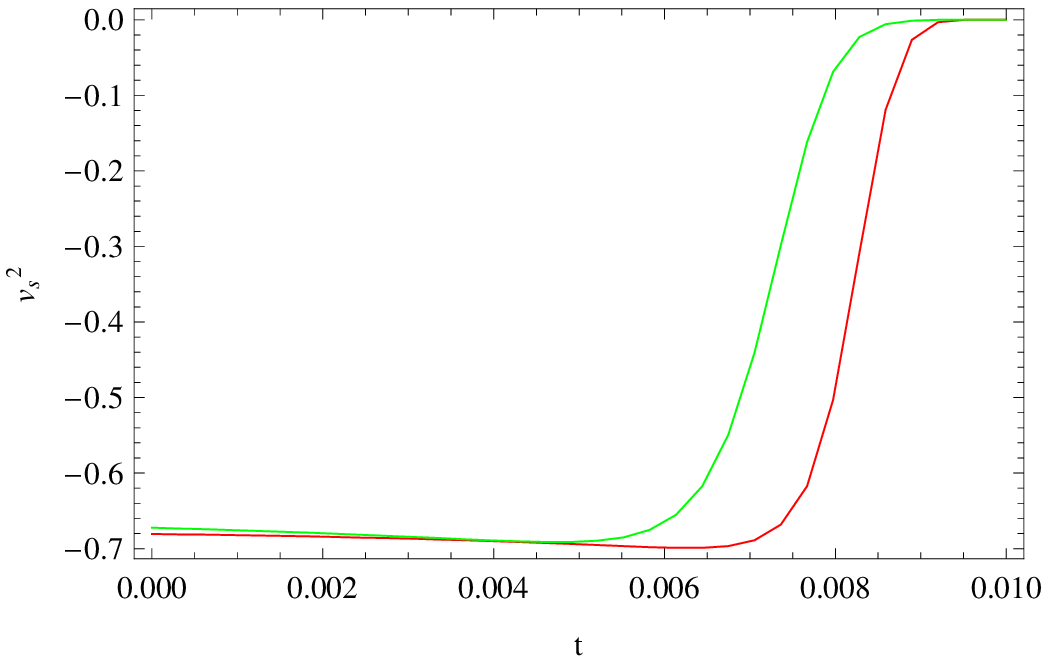}
\caption{\label{label}Evolution of the squared speed of sound $v_{s}^{2}$ given in Eq. (\ref{23}).}
\end{minipage}\hspace{3pc}%
\end{figure}

Using the expressions of the energy density $\rho_G$ and the pressure $p_G$ given, respectively, in Eqs. (\ref{19}) and (\ref{20}), we obtain that $v_{s}^{2}$ can be expressed as a function of cosmic time $t$ as follow:
\begin{equation}
v_{s}^{2}=\frac{\zeta_{1}}{\zeta_{2}}, \label{23}
\end{equation}
where:
\begin{eqnarray}
  \zeta_{1} &=& n a_{0}^{3}(t_{s}-t)^{5+3n}\left[-12\xi^{2}a_{0}^{4}(n-1)^{5}(t_{s}-t)^{-5+4n} \right. \nonumber \\
 && \left. -6a_{0}^{2}(n-1)^{3}n^{2}(t_{s}-t)^{-3+2n}m_{p}^{2}\right], \\  \label{21-1}
\zeta_{2} &=&4a_{0}^{7}(n-1)^{5}(7n-4)(t_{s}-t)^{7n}\xi^{2} \nonumber \\
&&+3n^{2}(t_{s}-t)^{4}\left(a_{0}(t_{s}-t)^{n}\right)^{-3\delta}\delta^{2}\rho_{m0} \nonumber \\
&& +2a_{0}^{5}(n-1)^{3}n^{2}(5n-2)(t_{s}-t)^{2+5n}m_{p}^{2}.\label{22-2}
\end{eqnarray}
The behavior of $v_{s}^{2}$ is plotted in Figure 2.
It is clear from Figure 2 that $v_s^2$ stays at negative
level for the DE based on GUP with the choice of scale factor made in this paper. Thus, we can conclude that this DE
is unstable for the expression of scale factor considered.

\section{Concluding remarks}
In this work, we studied the characteristics of NADE model in the
light of the GUP. We derived the EoS parameter $\omega_G$ given in
Eq. (\ref{21}) and we plotted it in Figure 1, which has
shown a phantom-like behavior of $\omega_G$. Moreover, we
also studied the main characteristics of the squared speed of
sound $v_s^2$, which expression, given in Eq. (\ref{23}), is
plotted in Figure 2. This Figure proves the instability of the DE
based on GUP with power law form of the scale factor considered in this work.
\\\\
\subsection{Acknowledgement}
The second author (SC) acknowledges the research grant under Fast
Track Programme for Young Scientists provided by the Department of
Science and Technology (DST), Govt of India. The project number is
SR/FTP/PS-167/2011.
\\

\end{document}